\documentclass[%
preprint,
 amsmath,amssymb,
 aps,
]{revtex4-1}
\usepackage{graphicx}
\graphicspath{{./Figures/}}
\usepackage{slashed}
\usepackage{booktabs}
\usepackage{graphicx}%
\usepackage{dcolumn}%
\usepackage{bm}%
\usepackage{hyperref}%

\begin{document}

\title{Determination of $f_0 - a_0^0$ mixing angle from QCD sum rules}

\author{T.M.Aliev}
  \email{taliev@metu.edu.tr}
 \author{S. Bilmis}%
\affiliation{%
Department of Physics, Middle East Technical University, 06800, Ankara, Turkey\\
}%

\date{\today}%

\begin{abstract}
By assuming that the $f_0$ and $a_0^0$ mesons are mixed states of the  two-quark - tetraquark, the mixing angle between them is estimated within QCD sum rules method, and it is obtained that the mixing angle is $(6.03 \pm 0.08)^\circ$. Our prediction on mixing angle can be checked in further experiments which can shed light on choosing the ``right'' structure of $f_0$ and $a_0^0$ mesons.
\end{abstract}

\maketitle

\section{\label{sec:intro}Introduction}
Although the light scalar mesons $a_0^0 (980)$ and $f_0(980)$ were discovered a long time ago, explanations of the nature of these states are still under debate. These states with similar mass but different isospins and decay channels are very problematic to accommodate in the quark model~\cite{Achasov:1987ts}, which created many discussions about the structure of these states. For example, tetraquarks~\cite{Achasov:1987ts,Weinstein:1983gd}, $K\bar{K}$ molecule~\cite{Weinstein:1990gu} or quark-antiquarks gluon hybrid state~\cite{Ishida:1995km}. The scalar mesons as a tetraquark state are intensively investigated within QCD sum rules in many studies (see~\cite{Wang:2015uha} and references therein). The main conclusion of these works is that the tetraquark picture described well existing many experimental data, but for an explanation of few data required modification of this picture. For example in~\cite{Sugiyama:2007sg,Kojo:2008hk} it is assumed that the scalar nonet represents as the superposition of tetraquarks and $\bar{q}q$ components. This picture is also necessary in order to eliminate the instanton effects~\cite{Hooft:2008we}. The mixing of these states firstly proposed in~\cite{Achasov:1979xc} (see also \cite{Wu:2007jh,Wu:2008hx,Aceti:2012dj,Roca:2012cv,Sekihara:2014qxa,Wang:2016wpc,Sakai:2017iqs}) and extensively studied in many reactions, such as $\gamma p \rightarrow p \pi^0 \eta$~\cite{Kerbikov:2000pu}, $\pi^- p \rightarrow \pi^0 \eta n$~\cite{Kudryavtsev:2001ee}, $p n \rightarrow d \pi^0 \eta$~\cite{Kudryavtsev:2001ee,Kudryavtsev:2002uu}, $d d \rightarrow \alpha \pi^0 \eta$~\cite{Patrignani:2016xqp}, but experimental verification was absent up to recent time.

In addition to a $\bar{q}q$ and $q\bar{q}q\bar{q}$ pictures, isoscalar glueball degrees of freedom should be taken into account due to the existence of two scalar nonets below $2~\rm{GeV}$ hence, the scalar mesons can be mixture of these states

Recently BES III Collaboration has reported the first observation of $a_0^0(980)$ and $f_0(980)$ mixing~\cite{Ablikim:2018pik} (see also \cite{Ablikim:2010aa}). In this work, the mixing intensities $\xi_{fa}$ and $\xi_{af}$ (for their definitions see~\cite{Ablikim:2018pik}) are measured.

Inspired with this measurement, in the present work, we calculate the mixing angle between $a_0^0 (980)$ and $f_0(980)$ meson within the QCD sum rules by following the approach presented in \cite{Aliev:2010ra} by assuming that these mesons are represented by the mixture of $\bar{q}q~-\bar{q}q \bar{q}q$- glueball states. 

The paper is organized as follows. In section~\ref{sec:level2}, we derive the QCD sum rules for $a_0^0 (980)$ and $f_0(980)$ mixing angle. Section~\ref{sec:level3} is devoted to the analysis of the sum rules obtained in the previous section. This section also contains our conclusion.

\section{Determination of the $a_0^0 (980)$ and $f_0(980)$ mixing angle from QCD sum rules}
\label{sec:level2}

Before presenting the details of calculations of the $a_0^0 (980)$ and $f_0(980)$ mixing angle, few words about the quark structure of these mesons as two quark and tetraquark states are in order. In the ideal mixing limit, in quark model the structure of $a_0^0 (980)$ and $f_0(980)$ symbolically can be written as;
\begin{equation}
  \label{eq:1}
  \begin{split}
    a_0^0(980) &= \frac{u\bar{u} -d\bar{d}}{\sqrt{2}} \\
      f_0(980) &= s\bar{s}
    \end{split}
  \end{equation}

 These mesons as tetraquark states in the ideal mixing limit can be represented as
 \begin{equation}
   \label{eq:3}
   \begin{split}
     f_0(980) &= \frac{us\bar{u}\bar{s} + ds \bar{d}\bar{s}}{\sqrt{2}} \\
     a_0^0(980) &= \frac{us\bar{u}\bar{s} - ds \bar{d}\bar{s}}{\sqrt{2}}
   \end{split}
 \end{equation}

 As we already noted, we will assume that the $a_0^0(980)$ and $f_0(980)$ mesons are the superposition of two-quark - tetraquark and glueball states. In other words, interpolating currents for $f_0$ and $a_0^0$, which we deal with QCD sum rules, are linear combinations of two-quark, tetraquark, and glueball currents.
 \begin{equation}
   \label{eq:4}
   \begin{split}
     J_{f_0}^{(0)} &= \cos{\theta_{f_0}} J_{f_0}^{(4)} + \sin{\theta_{f_0}} J_{f_0}^{(2)} + B G_{\mu \nu} G^{\mu \nu}\\
     J_{a_0^0}^{(0)} &= \cos{\theta_{a_0^0}} J_{a_0^0}^{(4)} + \sin{\theta_{a_0^0}} J_{a_0^0}^{(2)} + B^\prime G_{\mu \nu} G^{\mu \nu}
   \end{split}
 \end{equation}
 where $J_{f_0,a_0^0}^{(4)}$ and $J_{f_0,a_0^0}^{(2)}$ are interpolating currents representing tetraquark and two-quark states respectively. These currents are;
 \begin{equation}
   \label{eq:5}
   \begin{split}
     J_{f_0 (a_0^0)}^{(4)} &= \frac{\epsilon^{ijk} \epsilon^{imn}}{\sqrt{2}} \bigg[ (u^{j^T} C \gamma_5 s^k) (\bar{u}^m \gamma_5 C {\bar{s}^{n^T}}) + (-) ({d^j}^T C \gamma_5 s^k) (d^m \gamma_5 C {\bar{s}_n}^T)\bigg] \\
     J_{f_0}^{(2)} &= \frac{1}{6\sqrt{2}} (\langle \bar{u} u \rangle + \langle \bar{d} d)) \bar{s} s \\
     J_{a_0^0}^{(2)} &= -\frac{1}{6 \sqrt{2}} \langle \bar{s} s \rangle (\bar{u}u - \bar{d}d)
   \end{split}
 \end{equation}
 where $i,j,k,m,n$ are color indices and $C$ is the charge conjugation operator, $G_{\mu \nu}$ is the gluon field strength tensor, $B$ and $B^\prime$ are arbitrary parameters with mass square dimension.
 
The mixing angles $\theta_{f_0}$ and $\theta_{a^0}$ within the QCD sum rules method favor the ideal two-quark - tetraquark mixing angles are estimated in~\cite{Wang:2015uha}: $ \theta_{f_0}^0 \simeq 72.6^\circ$, $\theta_{a_0^0}^0 \simeq 84.3^\circ$ which we will use in our next discussions. It is well known that in the exact $SU(3)$ limit all mesons must have the same mass and flavor. However, if this symmetry is violated due to the different mass of quarks, then the mass eigenstates in general case do not coincide with flavor eigenstates. For this reason, the mass eigenstates can be considered as the linear combination of the flavor eigenstates. Hence, interpolating currents of $f_0(980)$ and $a_0^0(980)$ mesons can be written as a linear combination of currents presented in eq.~\eqref{eq:4}, i.e
 \begin{equation}
   \label{eq:6}
   \begin{split}
     J_{f_0} &= \cos{\theta} J_{f_0}^{(0)} + \sin{\theta} J_{a_0^0}^{(0)} \\
     J_{a_0^0} &= -\sin{\theta} J_{f_0}^{(0)} + \cos{\theta} J_{a_0^0}^{(0)} 
   \end{split}
 \end{equation}

 Our primary goal of the present work is to determine the mixing angle $\theta$. For this aim, we consider the following correlation function,
 \begin{equation}
   \label{eq:2}
   \begin{split}
     \Pi(q) = i \int d^4 x e^{iqx } \langle 0 | T \{ J_{f_0}(x) \bar{J}_{a_0^0}(0) \} |0 \rangle
   \end{split}
 \end{equation}
 The currents $J_{f_0}$ and $J_{a_0^0}$ create from vacuum only $f_0$ and $a_0^0$ mesons respectively, and obviously, the phenomenological part of the correlation function should be equal to zero.

 Here we would like to make the following remark. Both the nearly-degenerate $a_0^0(980)$ and $f_0(980)$ mesons can decay into $K\bar{K}$. Due to the isospin breaking effect, the charged and neutral Kaon thresholds are different about $8~\rm{MeV}.$ In other words, $f_0(980)$ and $a_0(980)$ can also be described by KK molecule. In our next discussion, this possibility is not taken into account since in QCD sum rules it is very difficult (even may be impossible) to calculate the contribution of two particle states.

Using~\eqref{eq:6} from eq.~\eqref{eq:2} one can easily obtain that
\begin{equation}
   \label{eq:7}
   \begin{split}
     &- \sin{\theta} \cos{\theta}~ \big( \Pi_{f_0 f_0} + 2B \alpha_s \sin{\theta_0} \Pi_{f_0 g} + B^2   \Pi_{gg} \big) \\
     &+ \sin{\theta} \cos{\theta}~ \big( \Pi_{a_0^0 a_0^0} + 2B^\prime \alpha_s \sin{\theta_1} \Pi_{a_0^0 g} + B^{\prime2}   \Pi_{gg} \big)\\
     &+\cos^2{\theta} \big(\Pi_{f_0 a_0^0} + B^\prime \alpha_s \sin{\theta_0} \Pi_{f_0 g} + B \alpha_s \sin{\theta_1} \Pi_{a_0^0 g} + B B^\prime   \Pi_{gg} \big) \\
     &- \sin^2{\theta} \big(\Pi_{a_0^0 f_0} + B \alpha_s \sin{\theta_1} \Pi_{a_0^0 g} + B^\prime \alpha_s \sin{\theta_1} \Pi_{f_0 g} + B B^\prime   \Pi_{gg} \big) = 0
   \end{split}
 \end{equation}
 where,
 \begin{equation}
   \label{eq:12}
   \begin{split}
     \Pi_{ij} =  \int d^4 x e^{i q x} &\langle 0 | T \{j_i (x) \bar{j}_j(0)\} |0 \rangle \hspace{1cm} (i= f_0, a_0^0, g; ~ j = f_0, a_0^0, g)
   \end{split}
 \end{equation}
here we denote $g$ as a gluon field.
 
 The correlation function eq.\eqref{eq:2} can be calculated in terms of quarks and gluons in the deep Euclidean domain $q^2 << 0$, by using the operator product expansion (OPE).
 For calculation of this correlation function at the deep Euclidean domain, we insert the expressions of the interpolating currents given by eq.\eqref{eq:5} in eq.\eqref{eq:6} and contracting the quark fields in accordance with Wick's theorem. In result, one can obtain the correlation function in terms of light quark propagators. In subsequent calculations for the light quark propagator, we utilize the following form
 \begin{equation}
   \label{eq:8}
   \begin{split}
     S_q(x) &= \frac{i \slashed{x}}{2 \pi^2 x^4} - \frac{m_q}{4 \pi^2 x^2} - \frac{\langle \bar{q} q \rangle}{12} +i \frac{m_q \langle \bar{q} q \rangle \slashed{x}}{48} 
     - \frac{x^2}{192}m_0^2 \langle \bar{q} q \rangle \big[ 1 - \frac{i \slashed{x}m_q}{6} \big] \\ 
     &-i \frac{g G^{\alpha \beta}}{32 \pi^2 x^2} \big[ \slashed{x} \sigma_{\alpha \beta} + \sigma_{\alpha \beta} \slashed{x} \big]
     -\frac{1}{32 \pi^2}m_q \big[ \ln{\frac{-x^2 \Lambda^2}{4}} + 2\gamma_E \big] g_s G_{\alpha \beta} \sigma^{\alpha \beta}\\
     &- \frac{i x^2 \slashed{x} g^2 \langle \bar{q} q \rangle ^2}{7776} -\frac{x^4 \langle \bar{q} q \rangle \langle g^2 G^2 \rangle}{27648} + ...
   \end{split}
 \end{equation}
where $\Lambda$ is the parameter separating perturbative and non-perturbative domains and its values lies in the domain $\Lambda = (0.5 ; 1)$~\rm{GeV}~\cite{Chetyrkin:2007vm} and $G_{\alpha \beta}$ is the gluon field strength tensor.

 Performing the Fourier transformation one can find the result for the correlation function in momentum space. Furthermore performing the Borel transformation over variable $-q^2$, we get desired sum rules. An important step to improve the sum rules is the subtraction procedure by using the hadron-quark duality approximation. According to duality approximation, the sum over excited states and continuum contributions are equal to the perturbative part of the correlation function starting from some $s_0$. After the subtraction procedure for the invariant functions $\Pi_{f_0 a_0^0}$, $\Pi_{f_0 f_0}$ and $\Pi_{a_0^0 a_0^0}$ we get
 \begin{equation}
   \label{eq:9}
   \begin{split}
     \Pi_{f_0 f_0} &= \Pi_{f_0f_0}^{22}  \sin^2{\theta_{f_0}} + 2 \Pi_{f_0f_0}^{42} \sin{\theta_{f_0}} \cos{\theta_{f_0}} + \Pi_{f_0f_0}^{44} \cos^2{\theta_{f_0}} \\
     \Pi_{a_0^0 a_0^0} &= \Pi_{a_0^0a_0^0}^{44}  \cos^2{\theta_{a_0^0}} + 2 \Pi_{a_0^0a_0^0}^{42} \sin{\theta_{a_0^0}} \cos{\theta_{a_0^0}} + \Pi_{a_0^0a_0^0}^{22} \sin^2{\theta_{a_0^0}}\\ 
     \Pi_{f_0 a_0^0} &= \Pi_{f_0a_0^0}^{44}  \cos{\theta_{f_0}} \cos{\theta_{a_0^0}} +  \Pi_{f_0a_0^0}^{42} \sin{\theta_{a_0^0}} \cos{\theta_{f_0}} + \Pi_{f_0a_0^0}^{24} \sin{\theta_{f_0}} \cos{\theta_{a_0^0}} + \Pi_{f_0a_0^0}^{22} \sin{\theta_{f_0}} \sin{\theta_{a_0^0}}
   \end{split}
 \end{equation}
 Expressions of the invariant functions  $\Pi_{f_0f_0}^{22;42;44}$, $\Pi_{a_0^0 a_0^0}^{22;42;44}$  and $\Pi^{4}_{f_0(a_0^0) g}$ functions are lengthy, and for this reason we do not present them here. It should be noted that, $\Pi_{f_0f_0}^{22;42;44}$ and $\Pi_{a_0^0 a_0^0}^{22;42;44}$ in the $SU(2)$ symmetry limit was derived in~\cite{Wang:2015uha}. In $SU(2)$ symmetry violation case, we recalculated these invariant functions.
 
 The expressions of the correlation functions $\Pi_{f^0 g}$, $\Pi_{a_0^0g}$ and $\Pi_{gg}$ are (see \cite{HARNETT2011110,Zhang:2009qb}).

 \begin{equation}
   \label{eq:13}
   \begin{split}
     \Pi_{f_0 g} &=\frac{-1}{6 \sqrt{2}} (\langle \bar{u}u \rangle + \langle \bar{d}d \rangle) \bigg( -\frac{23}{2 \pi} (\frac{\alpha_s}{\pi})^2 m_s M^4 + \frac{3}{2 \pi} (\frac{\alpha_s}{\pi})^2 2 m_s (M^4 \ln{M^2} -M^4 \gamma_E +M^4) \\
     &+ \langle \bar{s} s \rangle 8 \pi (\frac{\alpha_s}{\pi})^2 M^2 
     + m_s \langle \alpha_s G^2 \rangle (6 \frac{\alpha_s}{\pi} -2\frac{\alpha_s}{\pi} (\ln{M^2} - \gamma_E)) + 4\alpha_s m_0^2  \langle \bar{s} s \rangle \bigg),\\
     \\
 \Pi_{a_0 g} &= \frac{-1}{6 \sqrt{2}} \langle \bar{s} s  \rangle  \bigg\{   \bigg( -\frac{23}{2 \pi} (\frac{\alpha_s}{\pi})^2 m_u M^4 + \frac{3}{2 \pi} (\frac{\alpha_s}{\pi})^2 2 m_u (M^4 \ln{M^2} -M^4 \gamma_E +M^4) \\
     &+ \langle \bar{u} u \rangle 8 \pi (\frac{\alpha_s}{\pi})^2 M^2 
     + m_u \langle \alpha_s G^2 \rangle (6 \frac{\alpha_s}{\pi} -2\frac{\alpha_s}{\pi} (\ln{M^2} - \gamma_E)) + 4\alpha_s m_0^2  \langle \bar{u} u \rangle \bigg) \\
     &- \bigg( -\frac{23}{2 \pi} (\frac{\alpha_s}{\pi})^2 m_d M^2 + \frac{3}{2 \pi} (\frac{\alpha_s}{\pi})^2 2 m_d (M^2 \ln{M^2} -M^2 \gamma_E + M^2) \\
     &+ \langle \bar{d} d \rangle 8 \pi (\frac{\alpha_s}{\pi})^2 M^2 
     + m_d \langle \alpha_s G^2 \rangle (6 \frac{\alpha_s}{\pi} -2 \frac{\alpha_s}{\pi} (\ln{M^2} - \gamma_E)) + 4\alpha_s m_0^2  \langle \bar{d} d \rangle \bigg)
     \bigg\}, \\
     \\
          \Pi_{gg} &= \int_0^s ds s^2 e^{-s/M^2} \bigg\{ 2(\frac{\alpha_s}{\pi})^2 \big[ 1+\frac{659}{36} \frac{\alpha_s}{\pi} +247.48 (\frac{\alpha_s}{\pi})^2 \big] \\
     &- 4(\frac{\alpha_s}{\pi})^3 \big(\frac{9}{4}+65.781 \frac{\alpha_s}{\pi} \big) \ln \frac{s}{\mu^2} -10.125 (\frac{\alpha_s}{\pi})^4 (\pi^2 - 3 \ln^2 \frac{s}{\mu^2}) \bigg\} \\
     &+ 9\pi (\frac{\alpha_s}{\pi})^2 \langle \alpha_s G^2 \rangle \int_0^{s_0} ds e^{-s/M^2} + 8\pi^2 (\frac{\alpha_s}{\pi})^2 \langle \mathcal{G}_6 \rangle - 8 \pi^2 \frac{\alpha_s}{\pi} \frac{1}{M^2} \langle \mathcal{G}_8 \rangle 
   \end{split}
 \end{equation}
where
 \begin{equation}
   \label{eq:17}
   \begin{split}
     \langle \mathcal{G}_6  \rangle =& \langle g_s f_{abc} G_{\mu \nu}^a G_{\nu \rho}^b G_{\rho \mu}^c \rangle = (0.27~\rm{GeV^2}) \langle \alpha_s G^2 \rangle \\ 
     \langle \mathcal{G}_8  \rangle =& \langle ( \alpha_s  f_{abc} G_{\mu \nu}^a G_{\nu \rho}^b )^2 \rangle - \langle( \alpha_s  f_{abc} G_{\mu \nu}^a G_{\rho \lambda}^b )^2 \rangle \\
     =& \frac{9}{16} (\langle \alpha_s G^2 \rangle)^2~.
   \end{split}
 \end{equation}

 \begin{figure}[h]
  \centering
  \includegraphics[scale=0.5]{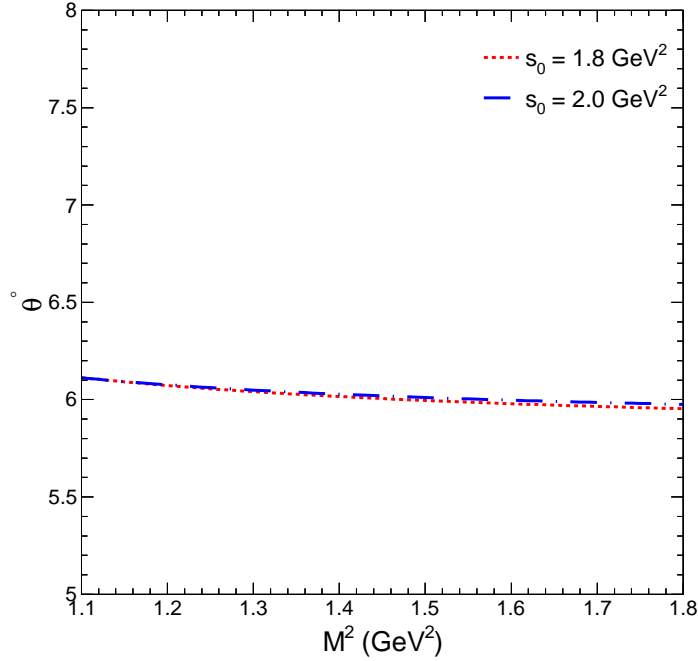}
  \caption{The dependence on mixing angle $\theta$ on Borel mass parameter square $M^2$ at two fixed values of $s_0$; $s_0 = 1.8~\rm{GeV^2}$ and $s_0 = 2.0~\rm{GeV^2}$. The upper(lower) line corresponds to the case when the glueball components are taken into account (without glueball component).}
  \label{fig:1}
\end{figure}

 \section{\label{sec:level3} Numerical Analysis}
 In this section, we perform numerical analysis of the sum rule for the mixing angle $\theta$. From eqs.~\eqref{eq:7} and \eqref{eq:9}, it follows that to perform the numerical analysis we need some input parameters. The values of the input parameters are presented in Table~\ref{tab:values}. The parameter $\gamma$ characterized the isospin breaking effects in quark condensates. Its value is calculated in~\cite{Jin:1994jz} and it is found that $-3 \times 10^{-3} \leq \gamma \leq 10^{-2}$. The values of masses are presented in the $\overline{MS}$ scheme. In calculations, we used the quark masses at $1~\rm{GeV}$.

 \begin{table}[hbt]
   \renewcommand{\arraystretch}{1.2}
   \setlength{\tabcolsep}{10pt}
   \centering
   \begin{tabular}{lc}  
\toprule
  Parameters & Values \\
  \hline
\midrule
$\overline{m_u}$ (\rm{2 GeV})      &
                                     $2.2^{+0.6}_{-0.4}~\rm{MeV}$ 
     \\
$\overline{m_d}$ (\rm{2 GeV})         &
                                        $4.7^{+0.5}_{-0.4}~\rm{MeV}$ 
     \\
$\overline{m_s}$ (\rm{2 GeV})       &
                                      $96^{+8}_{-4}~\rm{MeV}$
     \\
$\langle \bar{u} u \rangle$ (\rm{1 GeV})      & $ (-0.246^{+28}_{-19})^3$ $\rm{GeV^3}$         \\
  $\langle \bar{d} d \rangle$ (\rm{1 GeV}) & $(1 + \gamma) \langle \bar{u} u \rangle $           \\
     $ m_0^2$ & ($0.8 \pm 0.1$) $\rm{GeV^2}$ \\
     $\langle \frac{\alpha_s}{\pi} G^2 \rangle$ & $ (0.012 \pm 0.006)~\rm{GeV^4}$\\
  \bottomrule
 \hline
\end{tabular}
   \caption{Numerical values of the input parameters used in the calculations.}
   \label{tab:values}
 \end{table}

 Few words about the parameter $B$ and $B^\prime$. In principle, these parameters can be obtained from the analysis of two-point sum rules. In \cite{Wang:2015uha}, it is obtained that the masses of $f_0(980)$ and $a_0^0(980)$ are well reproduced without the gluon components in the current if these mesons are represented as a mixture of tetraquark and two-quark states. Moreover, the gluonic components of $a_0(1450)$ and $f_0(1370)$ states are also negligible \cite{Zhang:2009qb}. For these reasons in performing numerical analysis, we will set $B = B^\prime = 0$. If the future experiments indicate the existence of gluonic components in $f_0(980)$ and $a_0^0(980)$ states, then the afore-presented results can be used for the relevant analysis.

In addition to these input parameters, the sum rules contain two additionally auxiliary parameters: Borel mass parameter $M^2$ and the continuum threshold $s_0$. The mixing angle between $f_0$ and $a_0^0$ obviously should be independent on these parameters. Generally, the continuum threshold $s_0$ is not arbitrary and it is related to the first excited state energy as $s_0 = (m_{\text{ground}} + \delta)^2$ where $\delta$ is the energy difference between first and ground state energy. Analysis of various sum rules predict that $\delta$ lies between ($0.3~\rm{GeV}$ to $0.9~\rm{GeV}$.) For more precise determination of $s_0$ we impose the dominance of pole contribution and OPE convergence conditions. In result, we get $1.8~\rm{GeV^2} \leq s_0 \leq 2.6~\rm{GeV^2}$.

 For determination of working region of $M^2$, following procedure has been used. The upper bound of $M^2$ is obtained by imposing the condition that the higher states and continuum contributions constitute less than say $40\%$ of the total result. In order to find lower bound of $M^2$, we demand that the contributions of higher dimensional operators less than $25\%$ of the total results. These two conditions lead to the following working domain of $M^2$.
 \begin{equation}
   \label{eq:10}
   1.0~\rm{GeV^2} \leq M^2 \leq 1.8~\rm{GeV^2}
 \end{equation}

 In Fig.\ref{fig:1},
 we present the dependence of the mixing angle $\theta$ on $M^2$ at two fixed values of $s_0 = 1.8~\rm{GeV^2}~\text{and } 2~\rm{GeV^2}$. From this figure, it follows that the mixing angle $\theta$ exhibits good stability to the variation of $M^2$ or $s_0$. Taking into account the uncertainties of the input parameters as well uncertainties coming from the variation of $M^2$ and $s_0$ for mixing angle we finally get
 \begin{equation}
   \label{eq:11}
   \theta = (6.03 \pm 0.08)^\circ .
 \end{equation}

 Finally, let compare our prediction on mixing angle with \rm{BES III} result. The mixing intensity  $\xi_{fa}$ is obtained in \cite{Ablikim:2018pik} as;
 \begin{equation}
   \label{eq:14}
   \begin{split}
     \xi_{fa} &= (0.99 \pm 0.16 \pm 0.30 \pm 0.09) \times 10^{-2} ~~~ \text{(solution-1)} \\
     \xi_{fa} &= (0.41 \pm 0.13 \pm 0.17 \pm 0.13) \times 10^{-2} ~~~ \text{(solution-2)}~.
   \end{split}
 \end{equation}
Using the relation between the mixing intensity and mixing angle $|\xi_{fa}| \simeq \tan^2\theta$ and aforementioned experimental results, we find the mixing angle as
 \begin{equation}
   \label{eq:15}
   \begin{split}
      \theta = (5.45 &\pm 1.65)^\circ \hfill~~~ \text{(for solution-1)} \\
      \theta = (3.02 &\pm 2.21)^\circ \hfill~~~ \text{(for solution-2)}.
   \end{split}
 \end{equation}

 Comparing these results with our prediction on mixing angle, we observe that our result is in good agreement with the solution-1 result of BES III.
 
Our concluding remark is about the instanton contribution. In performing numerical analysis, we take into account the instanton contributions (see also \cite{PhysRevD.79.114033}) and obtained that this contribution for the determination of the mixing angle is less than $1\%$ i.e. practically negligible.
 
In conclusion, inspired by the recent observation of $f_0 - a_0^0$ mixing at BES III, we calculate the mixing angle between these states. We estimate the mixing angle between $f_0$ and $a_0^0$
and find that it is
$\theta = (6.03 \pm 0.08)^\circ$. And using the \rm{BES III} data, we obtain that our prediction on mixing angle is in good agreement with experimental result of BES III (solution 1).  Our prediction on mixing angle can be checked at future experiments to be conducted at BES III and other accelerators that can help to get information about the structure of $f_0$ and $a_0$ mesons.
\section{Acknowledgements}
We thank  Wencheng Yan for useful discussions on the BES III results.

\bibliography{mybib}{}

\end{document}